\documentclass[letterpaper,english,preprint, superscriptaddress, nofootinbib]{revtex4}
\usepackage{lmodern}
\usepackage[T1]{fontenc}
\usepackage[latin9]{inputenc}
\usepackage{longtable}
\usepackage{color}
\usepackage{graphicx}
\usepackage{amssymb}
\usepackage{esint}

\makeatletter

\providecommand{\tabularnewline}{\\}

\makeatother

\usepackage{babel}

\begin{document}

\title{Effects of line-tying on resistive tearing instability in slab geometry}

\author{Yi-Min Huang}

\email{yimin.huang@unh.edu}

\affiliation{Center for Magnetic Self-Organization in Laboratory and Astrophysical
Plasmas, University of Wisconsin, Madison, Wisconsin 53706}

\affiliation{Department of Physics, University of Wisconsin, Madison, Wisconsin
53706}

\author{Ellen G. Zweibel}

\email{zweibel@astro.wisc.edu}

\affiliation{Center for Magnetic Self-Organization in Laboratory and Astrophysical
Plasmas, University of Wisconsin, Madison, Wisconsin 53706}

\affiliation{Department of Physics, University of Wisconsin, Madison, Wisconsin
53706}

\affiliation{Department of Astronomy, University of Wisconsin, Madison, Wisconsin
53706}

\begin{abstract}
The effects of line-tying on resistive tearing instability in slab
geometry is studied within the framework of reduced magnetohydrodynamics
(RMHD).\citep{KadomtsevP1974,Strauss1976} It is found that line-tying
has a stabilizing effect. The tearing mode is stabilized when the
system length $L$ is shorter than a critical length $L_{c}$, which
is independent of the resistivity $\eta$. When $L$ is not too much
longer than $L_{c}$, the growthrate $\gamma$ is proportional to
$\eta$ . When $L$ is sufficiently long, the tearing mode scaling
$\gamma\sim\eta^{3/5}$ is recovered. The transition from $\gamma\sim\eta$
to $\gamma\sim\eta^{3/5}$ occurs at a transition length $L_{t}\sim\eta^{-2/5}$. 
\end{abstract}
\maketitle

\section{Introduction}

The study of the effects of line-tying on magnetohydrodynamic (MHD)
instabilities has a long history. In the solar corona, the magnetic
field lines of solar coronal loops are deeply anchored (line-tied)
to the much denser photosphere. It has long been suggested that because
of the stabilizing effect of line-tying, the loops can remain stable
for a much longer period of time than expected from MHD instability
growth times without line-tying. Raadu was the first to analyze the
line-tying stabilization of ideal internal kink mode using the energy
principle. \citep{Raadu1972} Later, more detailed analyses have been
done for various resistive \citep{VelliH1986,VelliH1989,VelliHE1990,Hassam1990}
and ideal MHD instabilities.\citep{VelliHE1990a,ZweibelB1992} In
recent years, there is a reviving theoretical interest in line-tying,
\citep{Hegna2004,RyutovCP2004,RyutovFIAM2006,HuangZS2006,EvstatievDF2006,DelzannoEF2007,DelzannoF2008,SvidzinskiML2008}
partly due to progress from plasma laboratory experiments.\citep{BergersonFFHKSS2006,SunIDFL2008}

In this paper, we study the effects of line-tying on resistive tearing
instability in one of the simplest settings, the slab geometry. The
tearing instability in slab geometry is discussed in many textbooks.
\citep{GoldstonR1995,Sturrock1994} The equilibrium is a simple current
layer where the current is along the $z$ direction, and the current
density only depends on $x$. Associated with the current layer is
a magnetic field $B_{y}$ which changes direction across the current
layer. This configuration is ideally stable. With resistivity, the
magnetic field can reconnect through the resistive tearing instability.
To simplify the analysis, usually a strong guide field perpendicular
to the reconnecting field is assumed such that the incompressible
approximation may be justified. The configuration we consider here
is to place two perfectly conducting end plates perpendicular to the
guide field. Therefore, it is the guide field that is line-tied to
the end plates. Although this configuration is rather simple, to our
knowledge, it has not been studied previously. Velli and Hood have
studied the line-tied tearing mode in slab geometry;\citep{VelliH1989}
but in their case, the conducting end plates are perpendicular to
the reconnecting field. The model we study here may be regarded as
a prototype of more complicated resistive instabilities that arise
in coronal heating models based on Parker's scenario.\citep{Parker1972,DmitrukGD1998,DmitrukG1999,DmitrukGM2003,RappazzoDEV2006,RappazzoVED2007,RappazzoVED2008}

The rest of the paper is organized as follows. Sec. \ref{sec:Physical-Model-and}
sets up the physical model and the governing equations of the system.
Sec. \ref{sec:Analytical-Considerations} discusses some general properties
that can be deduced from the equations without actually solving them.
Sec. \ref{sec:Numerical-Method} gives a concise description of the
numerical methods. The numerical results are presented in Sec. \ref{sec:Results}.
We summarize and conclude in Sec. \ref{sec:Summary-and-Discussions}.

\section{Physical Model and Governing Equations\label{sec:Physical-Model-and}}

To simplify the problem, we make the following assumptions: (1) the
plasma density is a constant; (2) a strong, constant guide field along
the $z$ axis is present; and (3) the length scales along the guide
field are much longer than the length scales in the perpendicular
directions. Under these assumptions, the system is may be described
by the well known reduced MHD (RMHD) equations. \citep{KadomtsevP1974,Strauss1976}
In this framework, after proper normalization, the magnetic field
and the velocity can be expressed in terms of the flux function $\psi$
and the stream function $\phi$ as $\mathbf{B}=\mathbf{\hat{z}}+\nabla_{\perp}\psi\times\mathbf{\hat{z}}$
and $\mathbf{u}=\nabla_{\perp}\phi\times\mathbf{\hat{z}}$. The governing
equations are 

\begin{equation}
\partial_{t}\Omega+\left[\phi,\Omega\right]=\partial_{z}J+\left[\psi,J\right]+\nu\nabla_{\perp}^{2}\Omega,\label{eq:RMHD1}\end{equation}
\begin{equation}
\partial_{t}\psi=\partial_{z}\phi+\left[\psi,\phi\right]+\eta\nabla_{\perp}^{2}\psi.\label{eq:RMHD2}\end{equation}
Where $\nu$ and $\eta$ are the viscosity and the resistivity; $\nabla_{\perp}\equiv\mathbf{\hat{x}}\partial_{x}+\mathbf{\hat{y}}\partial_{y}$
is the perpendicular gradient; $\Omega=-\nabla_{\perp}^{2}\phi$ ,
$J=-\nabla_{\perp}^{2}\psi$ are the vorticity and the electric current;
and $\left[\psi,\phi\right]\equiv\partial_{y}\psi\partial_{x}\phi-\partial_{x}\psi\partial_{y}\phi$
is the Poisson bracket. This set of RMHD equations is widely used
in studying the coronal heating and current sheet formation problems.\citep{LongcopeS1992,LongcopeS1992a,LongcopeS1992b,LongcopeS1994a,LongcopeS1994b,NgB1998,DmitrukGD1998,DmitrukG1999,DmitrukGM2003,RappazzoDEV2006,RappazzoVED2007,RappazzoVED2008}
Two perfectly conducting end plates are placed at $z=\pm L/2$ to
provide the line-tied boundary condition. \textcolor{black}{Although
the line-tied condition is an idealization in the sense that the solar
photosphere is neither completely rigid nor perfectly conducting,
it is a common approximation in solar physics. Strictly speaking,
the RMHD ordering $\partial_{z}\ll\nabla_{\perp}$ is incompatible
with the line-tied boundary condition, due to the presence of a boundary
layer near the end plate where $\partial_{z}\sim\nabla_{\perp}$.
However, Scheper and Hassam have shown through a boundary layer analysis
that RMHD can still be used in the coronal volume, as the boundary
layer only appears when the higher order correction to $B_{z}$ is
calculated. The higher order correction to $B_{z}$, however, is completely
decoupled from $\psi$ and $\phi$. }\citep{ScheperH1999}

We assume that the equilibrium only depends on $x$; i.e., $\mathbf{B}=\mathbf{\hat{z}}-\partial_{x}\psi\mathbf{\hat{y}}=\mathbf{\hat{z}}+B_{y}\mathbf{\hat{y}}$.
Consider perturbations of the form $\tilde{\phi}=\tilde{\phi}(x,z)\exp(ik_{y}y+\gamma t)$,
and $\tilde{\psi}=\tilde{\psi}(x,z)\exp(ik_{y}y+\gamma t)$. Linearizing
the RMHD equations gives \begin{equation}
\gamma\mathcal{D}\tilde{\phi}=(\partial_{z}+ik_{y}B_{y})\mathcal{D}\tilde{\psi}-ik_{y}B_{y}''\tilde{\psi}+\nu\mathcal{D}^{2}\tilde{\phi},\label{eq:linear1}\end{equation}
\begin{equation}
\gamma\tilde{\psi}=(\partial_{z}+ik_{y}B_{y})\tilde{\phi}+\eta\mathcal{D}\tilde{\psi}.\label{eq:linear2}\end{equation}
where $\mathcal{D}\equiv\partial_{x}^{2}-k_{y}^{2}$ and the prime
denotes $\partial_{x}$. The line-tied boundary condition in the RMHD
approximation is simply $\tilde{\phi}=0$ at $z=\pm L/2$. In this
study we will focus on the inertial tearing mode; hereafter the viscosity
$\nu=0$ is assumed throughout the rest of the paper.

\section{Analytical Considerations\label{sec:Analytical-Considerations}}

Although the linearized equations are difficult to solve analytically,
some general properties can be readily deduced from analytical considerations.
First of all, let us \emph{assume} that there exists a critical length
$L_{c}$ such that $\gamma=0$. By setting $\gamma=0$ and rescaling
$z$ with respect to $L_{c}$ (i.e., $z\rightarrow L_{c}z$), Eqs.
(\ref{eq:linear1}) and (\ref{eq:linear2}) become \begin{equation}
-\frac{1}{L_{c}}\partial_{z}\mathcal{D}\tilde{\psi}=ik_{y}B_{y}\mathcal{D}\tilde{\psi}-ik_{y}B_{y}''\tilde{\psi},\label{eq:marginal1}\end{equation}
\begin{equation}
-\frac{1}{L_{c}}\partial_{z}\tilde{\phi}=ik_{y}B_{y}\tilde{\phi}+\eta\mathcal{D}\tilde{\psi},\label{eq:marginal2}\end{equation}
with the boundary condition $\tilde{\phi}|_{z=\pm1/2}=0$. Eqs. (\ref{eq:marginal1})
and (\ref{eq:marginal2}) may be regarded as an eigenvalue problem
with $1/L_{c}$ as the eigenvalue. We further notice that by letting
$\tilde{\phi}\rightarrow\eta\tilde{\phi}$, the parameter $\eta$
completely drops out of the equations. Therefore, we conclude that
the critical length $L_{c}$ is independent of $\eta$. Some caveats
should be mentioned: (1) the existence of $L_{c}$ is not guaranteed;
(2) the marginal stability may not occur at $\gamma=0$, as pure oscillatory
modes are not precluded by the equations. Bearing these caveats in
mind, we found numerically that $L_{c}$ does exist, and indeed it
is independent of $\eta$. 

Now let us consider an unstable system not too far from marginality,
i.e., $L\gtrsim L_{c}$. If the unstable mode is so slowly growing,
we may neglect the inertia term (the left hand side) in Eq. (\ref{eq:linear1}).
Physically, that means that the system remains force-free throughout
the whole domain. Neglecting the inertia term in Eq. (\ref{eq:linear1}),
and dividing Eq. (\ref{eq:linear2}) by $\eta$, we come to the following
equations for the force-free approximation:\begin{equation}
0=(\partial_{z}+ik_{y}B_{y})\mathcal{D}\tilde{\psi}-ik_{y}B_{y}''\tilde{\psi},\label{eq:forcefree1}\end{equation}
\begin{equation}
(\gamma/\eta)\tilde{\psi}=(\partial_{z}+ik_{y}B_{y})(\tilde{\phi}/\eta)+\mathcal{D}\tilde{\psi}.\label{eq:forcefree2}\end{equation}
It immediately follows that in the force-free approximation, both
the growthrate $\gamma$ and the stream function $\tilde{\phi}$ scale
as $\eta$. Since the inertia term $\sim\gamma\tilde{\phi}\sim\eta^{2}$,
it is expected that the force-free approximation works better for
smaller $\eta$, when everything else remains the same. Recall that
in a periodic system, the magnetic field is approximately force-free
only in the {}``outer'' region, i.e., far from the resonant surface
where $\mathbf{k}\cdot\mathbf{B}=0$. In the inner layer, the resistivity
and the inertia are both taken into account. The inner layer thickness
$\delta\sim\eta^{2/5}$, and the growthrate $\gamma\sim\eta^{3/5}$
as a result.\citep{FurthKR1963,White1986} Naturally, we expect the
$\gamma\sim\eta^{3/5}$ scaling to be recovered if the system is sufficiently
long. The operative question is, at what length does this transition
occur? We will address this question later.

Finally, we would like to say a few words about the boundary condition.
Both Eq. (\ref{eq:linear1}) and Eq. (\ref{eq:linear2}) are first
order differential equations in $z$, for $\tilde{\psi}$ and $\tilde{\phi}$
respectively. Therefore two boundary conditions are needed. We have
two boundary conditions $\tilde{\phi}|_{z=\pm L/2}=0$ for $\tilde{\phi}$,
and none for $\tilde{\psi}$. If $\tilde{\psi}$ is given, then one
may integrate Eq. (\ref{eq:linear2}) along $z$ to solve $\tilde{\phi}$.
In general, the $\tilde{\phi}$ so obtained can not satisfy both boundary
conditions, unless some solubility condition for $\tilde{\psi}$ is
satisfied. Acting on (\ref{eq:linear2}) by $\int_{-L/2}^{L/2}dz\, e^{ik_{y}B_{y}z}$,
after integrating by parts and using the boundary condition on $\tilde{\phi}$,
we have the solubility condition \begin{equation}
\gamma\int_{-L/2}^{L/2}dz\, e^{ik_{y}B_{y}z}\tilde{\psi}=\eta\int_{-L/2}^{L/2}dz\, e^{ik_{y}B_{y}z}\mathcal{D}\tilde{\psi}.\label{eq:solubility}\end{equation}
This solubility condition is not used in the rest of the paper, but
we have used it to check the accuracy of the numerical solution.

\section{Numerical Method\label{sec:Numerical-Method}}

The numerical method we employ may be regarded as a variation of the
method proposed by Evstatiev \emph{et al}. in Ref. \citep{EvstatievDF2006}.
In essence, this method is a shooting method in two dimensions. We
may rewrite Eqs. (\ref{eq:linear1}) (\ref{eq:linear2}) in the following
form as an evolution equation along $z$: \begin{equation}
\frac{\partial}{\partial z}\left[\begin{array}{c}
\tilde{\psi}\\
\tilde{\phi}\end{array}\right]=\mathcal{L}_{\gamma}\left[\begin{array}{c}
\tilde{\psi}\\
\tilde{\phi}\end{array}\right],\label{eq:zeq}\end{equation}
where the operator $\mathcal{L}_{\gamma}$ is defined as ($\nu$ is
set to zero) \begin{equation}
\mathcal{L}_{\gamma}\equiv\left[\begin{array}{cc}
-ik_{y}\mathcal{D}^{-1}(B_{y}\mathcal{D}-B_{y}'') & \gamma\\
\gamma-\eta\mathcal{D} & -ik_{y}B_{y}\end{array}\right].\label{eq:L_gamma}\end{equation}
Starting from an initial condition $(\tilde{\psi}_{0}(x),\tilde{\phi}_{0}(x))$
at $z=0$, we can use Eq. (\ref{eq:zeq}) to propagate the initial
condition along $z$. Formally, the solution may be written as \begin{equation}
\left[\begin{array}{c}
\tilde{\psi}(x,z)\\
\tilde{\phi}(x,z)\end{array}\right]=\exp(\mathcal{L}_{\gamma}z)\left[\begin{array}{c}
\tilde{\psi}_{0}\\
\tilde{\phi}_{0}\end{array}\right].\label{eq:propagatation}\end{equation}
Apparently $\gamma$ is a eigenvalue if and only if there exists an
initial condition at $z=0$ such that when propagated to both end
caps by Eq. (\ref{eq:propagatation}), the boundary conditions are
satisfied. In the present case, we may consider the linear mapping
\begin{equation}
\mathcal{G}_{\gamma,L}:\left[\begin{array}{c}
\tilde{\psi}_{0}\\
\tilde{\phi}_{0}\end{array}\right]\mapsto\left[\begin{array}{c}
\tilde{\phi}(x,z=L/2)\\
\tilde{\phi}(x,z=-L/2)\end{array}\right].\label{eq:mapping}\end{equation}
To simplify the notation, hereafter we will use $\chi_{0}$ to denote
the initial condition. To satisfy the boundary conditions, we require
\begin{equation}
\sigma_{\gamma,L}\equiv\min_{\chi_{0}}\frac{\left|\left|\mathcal{G}_{\gamma,L}\chi_{0}\right|\right|}{\left|\left|\chi_{0}\right|\right|}=0,\label{eq:inf}\end{equation}
where $\left|\left|.\right|\right|$ is some suitable norm. Here we
use the straightforward generalization of the usual $L^{2}$ norm:\begin{equation}
\left|\left|\chi_{0}\right|\right|^{2}=\int_{-\infty}^{\infty}\left(\left|\tilde{\phi}_{0}\right|^{2}+\left|\tilde{\psi}_{0}\right|^{2}\right)dx,\label{eq:norm1}\end{equation}
\begin{equation}
\left|\left|\mathcal{G}_{\gamma,L}\chi_{0}\right|\right|^{2}=\int_{-\infty}^{\infty}\left(\left|\tilde{\phi}(x,L/2)\right|^{2}+\left|\tilde{\phi}(x,-L/2)\right|^{2}\right)dx.\label{eq:norm2}\end{equation}
Now if the system length $L$ is given, one can scan $\sigma_{\gamma,L}$
over the whole space of complex $\gamma$. Those $\gamma$'s where
$\sigma_{\gamma,L}=0$ are the eigenvalues. Alternatively one may
fix $\gamma$ and scan through $L$ to find out the system lengths
for which the given $\gamma$ is an eigenvalue. 

To implement this idea numerically, first we need an accurate discrete
approximation for the operator $\mathcal{L}_{\gamma}$. Pseudospectral
methods are suitable for the task.\citep{Trefethen2000,WeidemanR2000,Fornberg1995}
The numerical integrations for Eqs. (\ref{eq:norm1}) and (\ref{eq:norm2})
are done with Gaussian quadrature.\citep{PressTVF2002,TrefethenB1997}
Since we are dealing with an infinite domain as most textbooks do,
we need to map $(-\infty,\infty)$ smoothly to $(-1,1)$ to apply
pseudospectral methods. We define the mapping as follows. \citep{Boyd2001}
First 

\begin{equation}
x=\frac{c_{1}x_{1}}{\sqrt{1-x_{1}^{2}}}\label{eq:map1}\end{equation}
 maps $x_{1}\in(-1,1)$ to $x\in(-\infty,\infty)$. Then a second
mapping \begin{equation}
x_{2}=(1-c_{3})x_{1}+c_{3}\frac{2}{\pi}\tan^{-1}\left(\tan\left(\pi x_{1}/2\right)/c_{2}\right)\label{eq:map2}\end{equation}
maps $x_{1}\in(-1,1)$ to $x_{2}\in(-1,1).$ This mapping can further
concentrate resolution around $x=0$ to resolve the resistive layer.
The constants $c_{1}$, $c_{2}$, $c_{3}$ are free parameters. A
set of $N$ Legendre-Gauss collocation points\citep{Fornberg1995}
are set up in $x_{2}$. The differentiation matrices and Gaussian
quadrature weights are first calculated in $x_{2}$ with respect to
these nodes, then transformed back to $x$. In this discrete approximation,
the initial condition $\chi$ is represented by a column vector $q$,
which consists of $\tilde{\phi}_{0}$ and $\tilde{\psi}_{0}$ on the
collocation points. The operators $\mathcal{L}_{\gamma}$ and $\mathcal{G}_{\gamma,L}$
are represented by matrices $L_{\gamma}$ and $G_{\gamma,L}$, respectively.
The norm is represented by\begin{equation}
\left|\left|q\right|\right|^{2}=q^{\dagger}Wq,\label{eq:norm3}\end{equation}
where $\dagger$ denotes the conjugate transpose and $W$ is a real
diagonal matrix from the Gaussian quadrature. We have to find \begin{eqnarray}
\sigma_{\gamma,L} & = & \min_{q}\frac{(q^{\dagger}G_{\gamma,L}^{\dagger}WG_{\gamma,L}q)^{1/2}}{(q^{\dagger}Wq)^{1/2}}\label{eq:inf1}\end{eqnarray}
and the corresponding $q$ that gives the minimum. Let $W=M^{\dagger}M$
and $s=Mq$. Then\begin{equation}
\sigma_{\gamma,L}=\min_{s}\frac{(s^{\dagger}M^{-1\dagger}G_{\gamma,L}^{\dagger}M^{\dagger}MG_{\gamma,L}M^{-1}s)^{1/2}}{(s^{\dagger}s)^{1/2}}.\label{eq:inf2}\end{equation}
Now we calculate the singular value decomposition (SVD) \citep{TrefethenB1997}
$MG_{\gamma,L}M^{-1}=USV^{\dagger}$, where $U$ and $V$ are unitary
matrices of left and right singular vectors, and $S$ is a real nonnegative
diagonal matrix of singular values. Then $\sigma_{\gamma,L}$ is the
smallest singular value, and $q=M^{-1}s$, where $s$ is the right
singular vector (a column of $V$) corresponding to the smallest singular
value. The eigenfunction can be reconstructed by propagating $q$
along $z$ by $\exp(\mathcal{L}_{\gamma}z)$.

\begin{figure}
\begin{centering}
\includegraphics[scale=0.45]{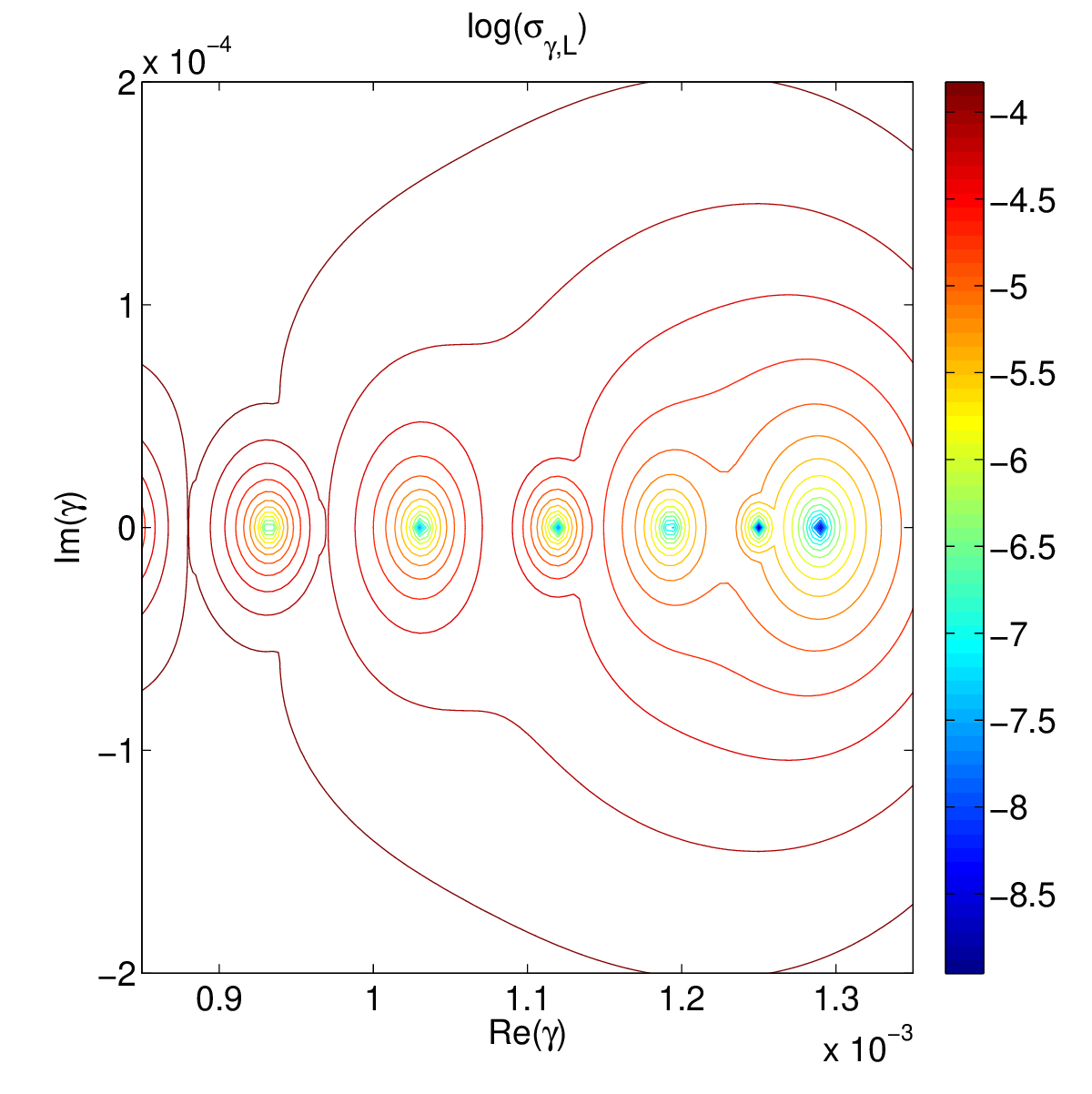}
\par\end{centering}

\caption{(Color online) $\log(\sigma_{\gamma,L})$ in a domain of complex $\gamma$.
$\eta=10^{-4}$ and $L=1000$. Those $\gamma$'s where $\sigma_{\gamma,L}$
goes to zero are eigenvalues. Here we found six eigenvalues within
this domain. All of them are on the real axis. \label{fig:scan2d} }

\end{figure}

Fig. \ref{fig:scan2d} shows $\log(\sigma_{\gamma,L})$ within a domain
of complex $\gamma$ for the case $\eta=10^{-4}$, $L=1000$. Those
$\gamma$'s where $\sigma_{\gamma,L}$ goes to zero are eigenvalues.
In this particular case, we found six eigenvalues within the domain,
all on the real axis. It is rather computationally expensive to scan
through a complex domain. Fortunately, we found that the most unstable
mode always has a real eigenvalue. Therefore we only scan through
the real axis for most of our calculation. 

Here we briefly compare our method with the method of Evstatiev \textit{et
al}. in Ref. \citep{EvstatievDF2006}. These authors let $\partial_{z}\rightarrow\lambda$
(they use $ik$ instead of $\lambda$) in Eq. (\ref{eq:zeq}) then
solve it as an auxiliary eigenvalue problem. Let $\lambda_{i}e_{i}=\mathcal{L}_{\gamma}e_{i}$,
where $\lambda_{i}$ is the eigenvalue and $e_{i}$ the eigenfunction
of $\mathcal{L}_{\gamma}$. The general solution is written as $\chi=\sum_{i=1}^{n}a_{i}e^{\lambda_{i}z}e_{i}$
(truncated to a finite number of $n$ modes), which satisfies Eq.
(\ref{eq:zeq}). To impose the boundary conditions, they consider
\begin{equation}
\bar{\sigma}_{\gamma,L}=\min_{a_{i}}\frac{\left(\sum_{j=1}^{m}\left|\chi_{j}\right|^{2}\right)^{1/2}}{\left(\sum_{i=1}^{n}\left|a_{i}\right|^{2}\right)^{1/2}},\label{eq:sigma1}\end{equation}
where we use $\chi_{j}$ to symbolically denote the value of the solution
(in the present case, the $\tilde{\phi}$ part only) on a set of $m$
points (which can be arbitrarily chosen, with $m\ge n$) on the end
plates. Scanning through $\gamma$ for a given $L$ (or through $L$
for a given $\gamma$), the local minimum of $\bar{\sigma}_{\gamma,L}$
corresponds to an eigenmode of the original eigenvalue problem. Now
it is clear that the two approaches are rather similar, and the differences
are only minor details. If the eigenmodes of $\mathcal{L}_{\gamma}$
form a complete basis, projecting onto the eigenbasis is a convenient
and efficient way of finding $\exp(\mathcal{L}_{\gamma}z)$. Explicitly,
if $\chi_{0}=\sum_{i}a_{i}e_{i},$ then $\exp\left(\mathcal{L}_{\gamma}z\right)\chi_{0}=\sum_{i}a_{i}e^{\lambda_{i}z}e_{i}.$
Instead of representing $\chi_{0}$ by its value on collocation points,
one may use the expansion coefficients $\{a_{i}\}$. The linear mapping
$\mathcal{G}_{\gamma,L}$, the norm $\left|\left|.\right|\right|$,
and $\sigma_{\gamma,L}$ can be similarly defined in terms of $\{a_{i}\}$.
Formulated in this way, our approach is nearly identical to theirs.
The only real difference between between the two is that we use $\sigma_{\gamma,L}$
instead of $\bar{\sigma}_{\gamma,L}$. It is worth pointing out, however,
that even if the eigenmodes of $\mathcal{L}_{\gamma}$ are not complete,
$\exp\left(\mathcal{L}_{\gamma}z\right)$ may still exist. %
\footnote{As a simple example, consider $A=[3,-2;2,-1],$ which has only one
eigenvalue $1$ as a double root and only one corresponding eigenvector
$(1,1)$; therefore $A$ does not have a complete eigenbasis. However,
$\exp(Az)=e^{z}\left[1+2z,-2z;2z,1-2z\right]$ is well defined.%
} Therefore our method does \emph{not }rely on the existence of a complete
eigenbasis. 

\begin{figure}
\begin{centering}
\includegraphics[scale=0.45]{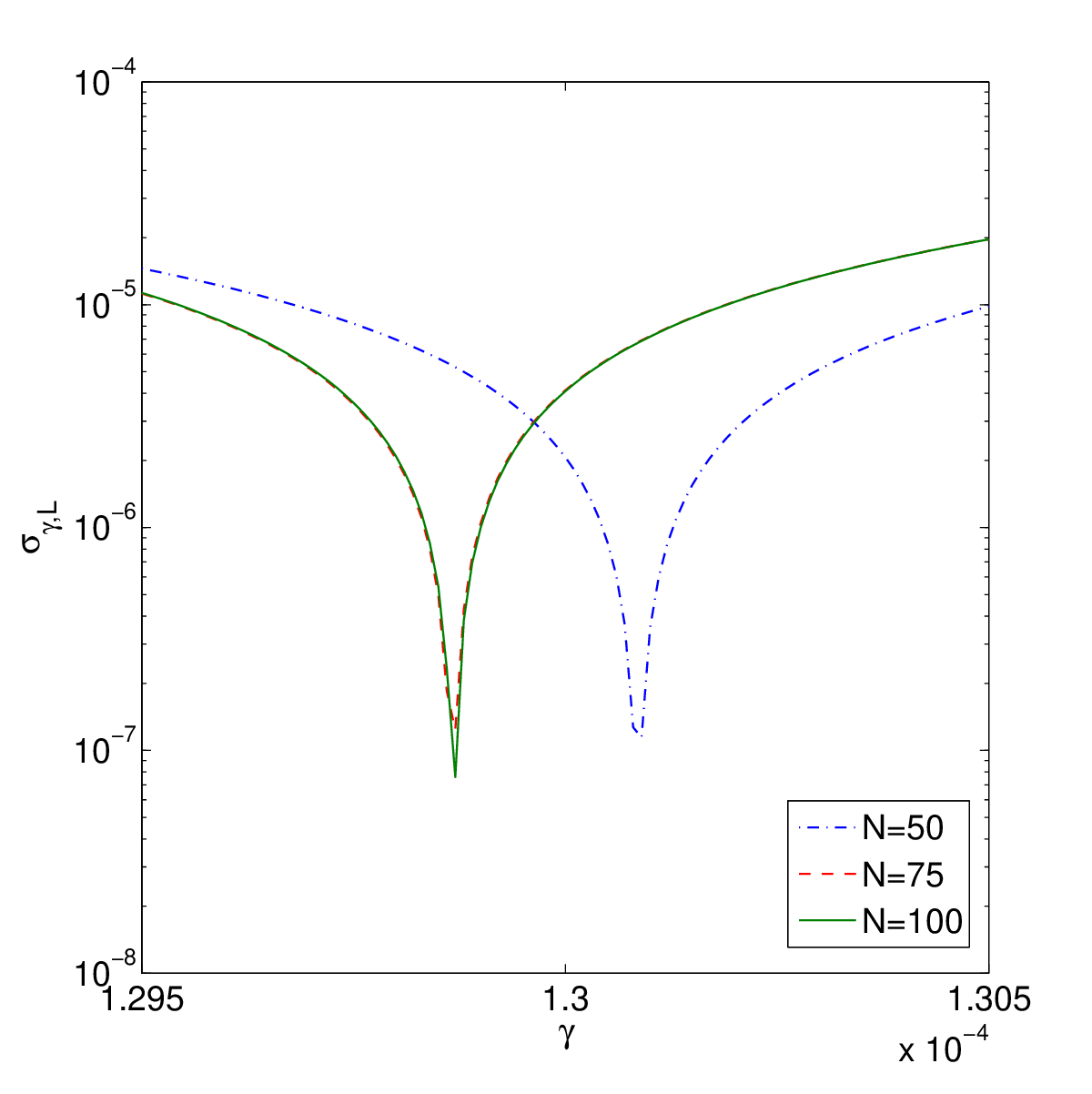}
\par\end{centering}

\caption{(Color online) Convergence test for the case $\eta=3\times10^{-6}$,
$L=2000$. The mapping parameters are $c_{1}=2$, $c_{2}=0.1$, $c_{3}=0.9$.
The $N=50$ calculation gives $\gamma\simeq1.301\times10^{-4}$, while
the other two calculations give $\gamma\simeq1.299\times10^{-4}$.
Note that the curves from $N=75$ and $N=100$ calculations are almost
identical, that indicates convergence. For $N>100$ calculations,
the results are virtually indistinguishable from the $N=100$ curve.
\label{fig:Convergence_Sigma}}

\end{figure}

\begin{figure}
\begin{centering}
\includegraphics[scale=0.5]{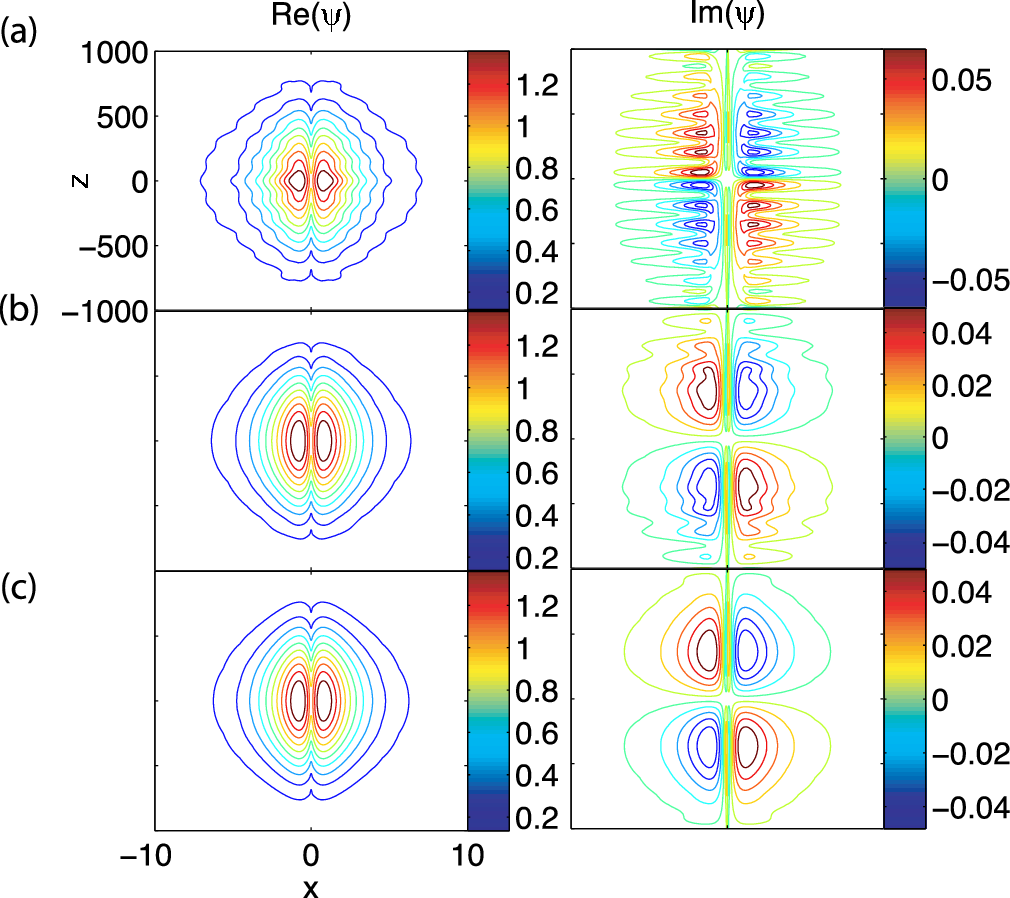}
\par\end{centering}

\caption{(Color online) The real and imaginary parts of eigenfunctions $\tilde{\psi}$
from the convergence test shown in Fig. \ref{fig:Convergence_Sigma}.
(a) N=50, (b) N=75, (c) N=100. Note that although the relative error
on eigenvalue from $N=50$ calculation is only about $0.2\%$, the
eigenfunction is clearly not accurate. The $N=75$ result is much
better, but there are still wiggles on the imaginary part. The $N=100$
calculation gives a fully converged eigenfunction.\label{fig:Convergence_eig}}

\end{figure}

The advantage of $\sigma_{\gamma,L}$ is that it has a precise, analytical
definition, Eq. (\ref{eq:inf}); and Eq. (\ref{eq:inf1}) is merely
a numerical approximation to it. This allows us an easier way to check
convergence --- if the numerical solution is fully converged, $\sigma_{\gamma,L}$
should be independent of the resolution and the mapping parameters.
On the other hand, the definition of $\bar{\sigma}_{\gamma,L}$ depends
on the points chosen to enforce the boundary condition, the set of
eigenmodes $\{e_{i}\}_{i=1}^{n}$ for expansion, as well as the normalization
of the eigenmodes. Fig. \ref{fig:Convergence_Sigma} shows a case
of our convergence test with respect to the resolution, for $N=50,$
$75,$ $100$. The $N=50$ calculation gives $\gamma\simeq1.301\times10^{-4}$,
while the other two calculations give $\gamma\simeq1.299\times10^{-4}$.
Note that the curves from $N=75$ and $N=100$ calculations are almost
identical, indicating convergence. For $N>100$ calculations, the
results are virtually indistinguishable from the $N=100$ curve. Fig.
\ref{fig:Convergence_eig} shows the eigenfunction $\tilde{\psi}$
from the same calculations. Although the relative error in the eigenvalue
from the $N=50$ calculation is only about $0.2\%$, the eigenfunction
is clearly not accurate. The $N=75$ result is much better, but there
are still wiggles on the imaginary part. The $N=100$ calculation
gives a fully converged eigenfunction. This once again demonstrates
that an accurate eigenfunction is much more difficult to obtain than
the eigenvalue. It took significant effort to make sure all the numerical
results presented in this work are fully converged.

\section{Results\label{sec:Results}}

We use the Harris sheet $B_{y}=b\tanh(x)$, $b=0.1$, as the equilibrium.
The wavenumber $k_{y}=0.5$ is used for most of the calculations.
The dependence on $k_{y}$ will be addressed later.

\subsection{Periodic System}

The periodic system is usually solved for the case $k_{z}=0$, but
the analysis can be easily generalized to $k_{z}\neq0$ cases. In
the so-called constant $\psi$ approximation (i.e. $\tilde{\psi}$
is approximately constant within the resistive layer), the growthrate
$\gamma$ is proportional to $\eta^{3/5}$.\citep{FurthKR1963,White1986}
Fig. \ref{fig:gamma_vs_kz} shows the growthrate $\gamma$ divided
by $\eta^{3/5}$ as a function of $k_{z}$, for different $\eta$.
In the asymptotic limit $\eta\rightarrow0$, all curves should lie
on top of each other. Some interesting features may be readily observed:
(1) Except for the case $\eta=10^{-4}$, $k_{z}=0$ is not the peak
of the growthrate. As we will see, this has an interesting consequence
in line-tied solutions. (2) The $\gamma\sim\eta^{3/5}$ scaling works
better around $k_{z}=0$. This may be due to the fact that it is easier
to satisfy the constant $\psi$ approximation there. However, we expect
that as long as $\eta$ is sufficiently small, eventually the constant
$\psi$ approximation will work for all $k_{z}$.

\textcolor{black}{At first sight, it may seem counter-intuitive that
$k_{z}=0$ is a local minimum of $\gamma$. In the tearing mode analysis
$\gamma$ is determined by $\Delta'$ and $(\mathbf{k}\cdot\mathbf{B})'$
at the resonant surface. For the Harris sheet profile $\Delta'$ may
be difficult to calculate analytically when $k_{z}\neq0$. However,
for a similar piecewise linear profile, i.e. $B_{y}=bx$ when $|x|\le1$
and $B_{y}=bx/|x|$ when $|x|\ge1$, an analytic $\Delta'$ can be
found. It may be shown that in this case $k_{z}=0$ is also a local
minimum. In fact, $\Delta'$ diverges as $k_{z}\rightarrow\pm k_{y}b$.
This pathological behavior may be due to the abrupt turning of $B_{y}$
at $x=\pm1$, which is unphysical. However, the result from this simple
system suggests that it may be generic, or at least rather unsurprising,
that $k_{z}=0$ is the local minimum for the system we consider here.}

\subsection{Line-tied System}

\begin{figure}
\begin{centering}
\includegraphics[scale=0.45]{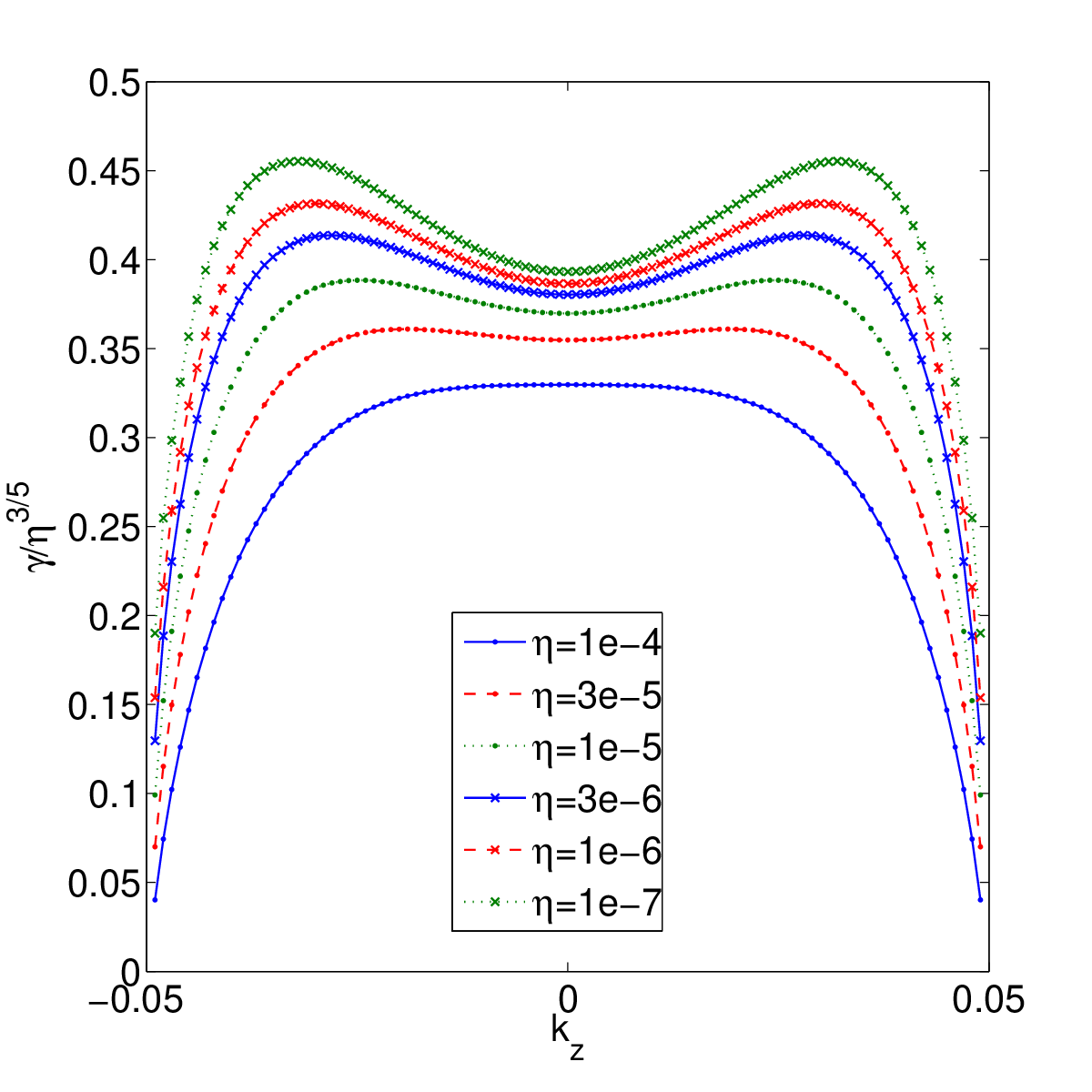}
\par\end{centering}

\caption{(Color online) $\gamma/\eta^{3/5}$ as a function of $k_{z}$, for
periodic systems. \label{fig:gamma_vs_kz}}

\end{figure}

\begin{figure}
\begin{centering}
\includegraphics[scale=0.45]{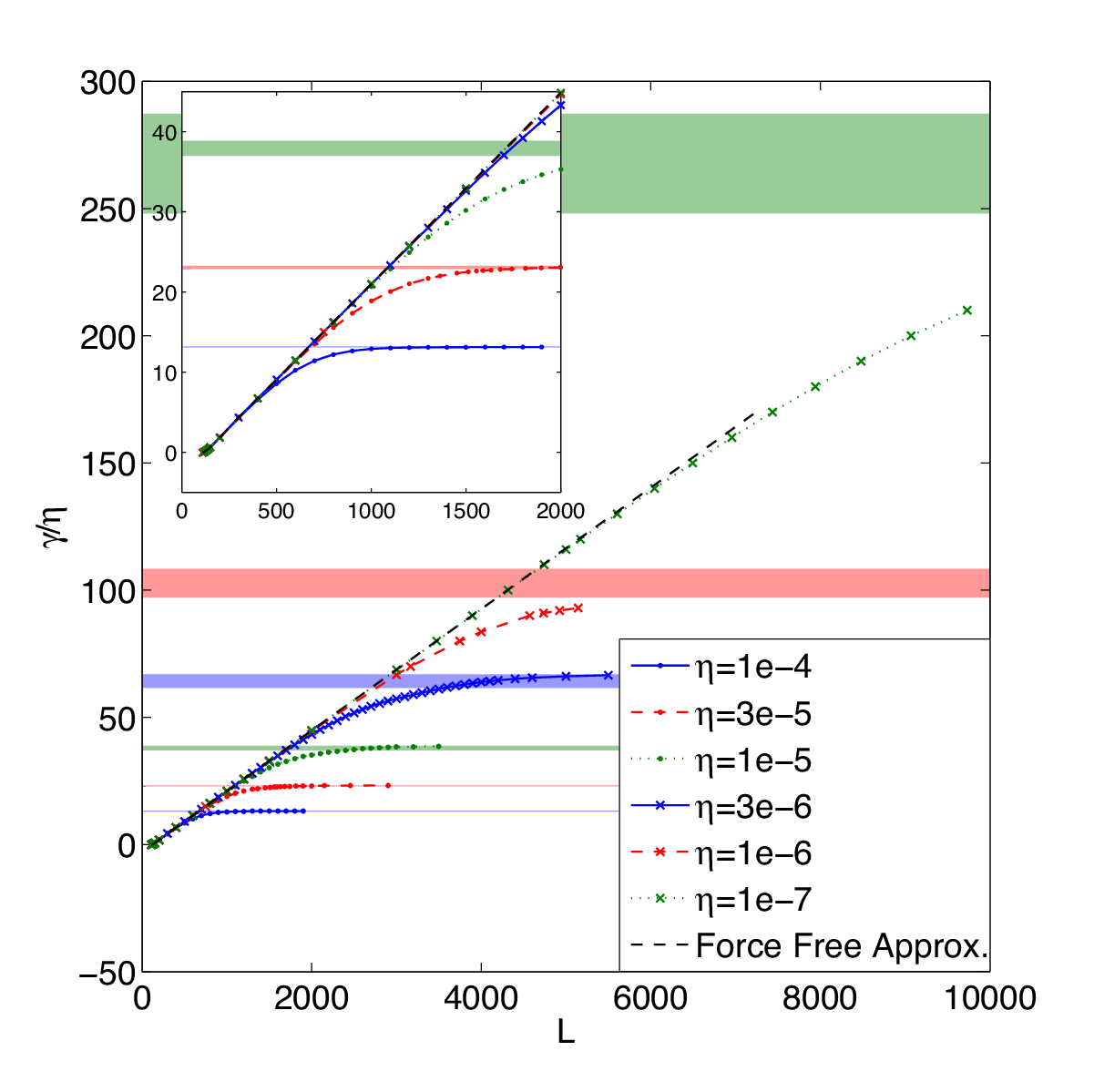}
\par\end{centering}

\caption{(Color online) $\gamma/\eta$ as a function of $L$, for various $\eta$.
The horizontal bars denote the range determined by the periodic $\gamma$
at $k_{z}=0$ and the the maximum periodic $\gamma$ (see Fig. \ref{fig:gamma_vs_kz}).
The dashed line is the prediction from the force-free approximation.
Inset: the expanded view of the lower left corner. \label{fig:gamma_vs_L}}

\end{figure}

The growthrate $\gamma$ of the fastest growing mode as a function
of $L$ for various $\eta$ is summarized in Fig. \ref{fig:gamma_vs_L}.
In the plot we normalized $\gamma$ with respect to $\eta$. As discussed
in Sec. \ref{sec:Analytical-Considerations}, we expect the critical
length $L_{c}$ to be independent of $\eta$, and $\gamma$ to be
proportional to $\eta$ near marginality. Both are borne out by the
numerical calculations. It is found that $L_{c}=115.09$ for all $\eta$,
and the $\gamma/\eta$ versus $L$ curves all coincide with the prediction
from the force-free approximation. As $L$ becomes longer, the curve
starts to deviate from the force-free approximation. The deviation
occurs earlier the larger $\eta$ is. This is also what we expect
in Sec. \ref{sec:Analytical-Considerations}. The horizontal bars
denote the range determined by the periodic $\gamma$ at $k_{z}=0$
and the maximum periodic $\gamma$ (see Fig. \ref{fig:gamma_vs_kz}).
We observe that as the growthrate peels away from the force-free approximation,
it first goes beyond the periodic $\gamma$ at $k_{z}=0$, and eventually
levels off at (but never exceeds) the maximum periodic $\gamma$.
Therefore the force-free approximation and the maximum periodic growthrate
may be regarded as two asymptotes of the line-tied growthrate. We
are not able to achieve fully converged results as the $\eta=10^{-6}$
and $10^{-7}$ cases approach their periodic limits. However, based
on the observation for larger $\eta$ cases, it is fairly safe to
say that the two-asymptote scheme may be generally applicable. 

\begin{figure}
\begin{centering}
\includegraphics[bb=210bp 0bp 640bp 779bp,clip,scale=0.55]{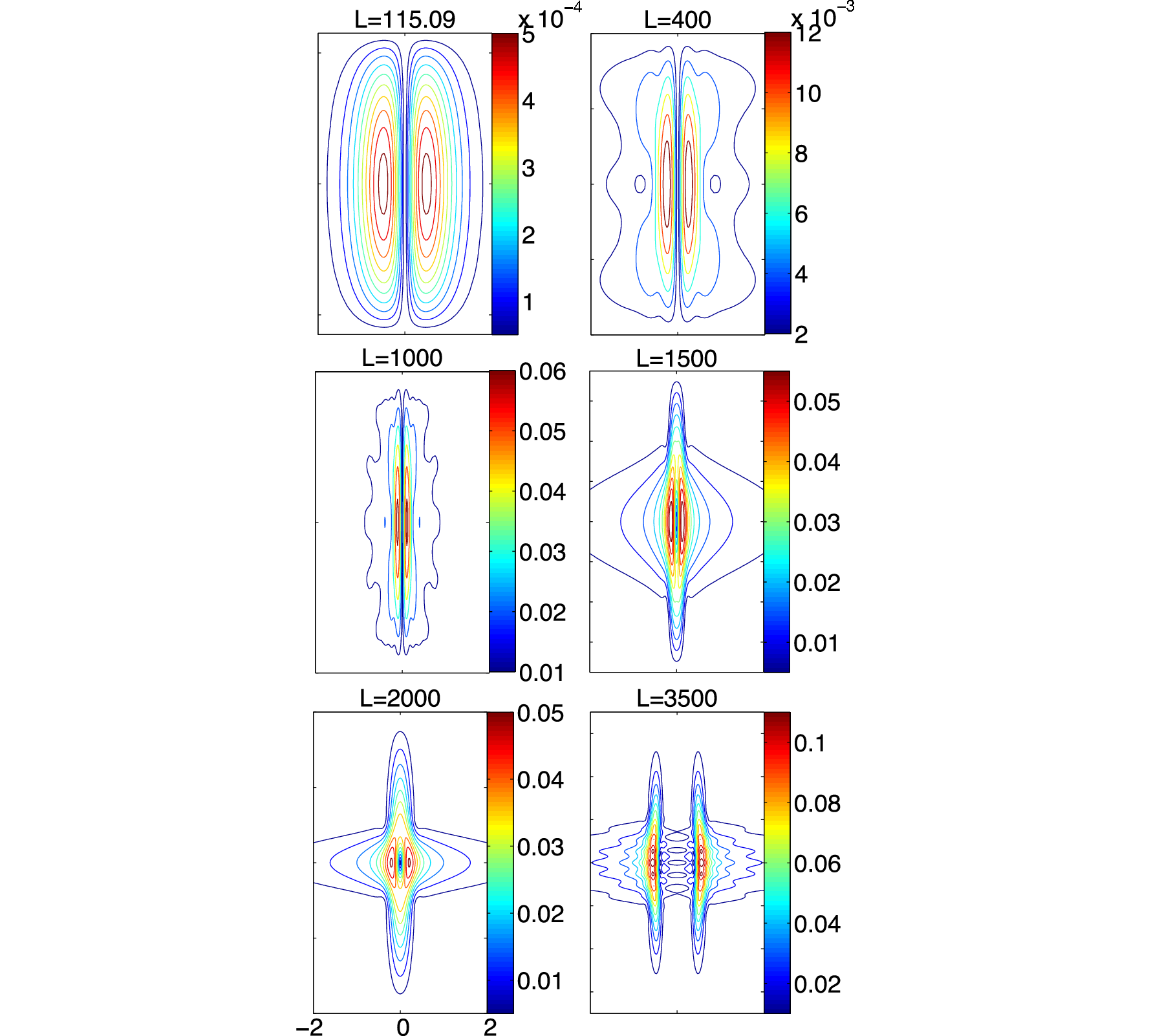}
\par\end{centering}

\caption{(Color online) The eigenfunction$\left|\tilde{\phi}\right|$ for various
$L$. $\eta=10^{-5}$.\label{fig:phi_eigen}}

\end{figure}

\begin{figure}
\begin{centering}
\includegraphics[bb=250bp 0bp 650bp 761bp,clip,scale=0.55]{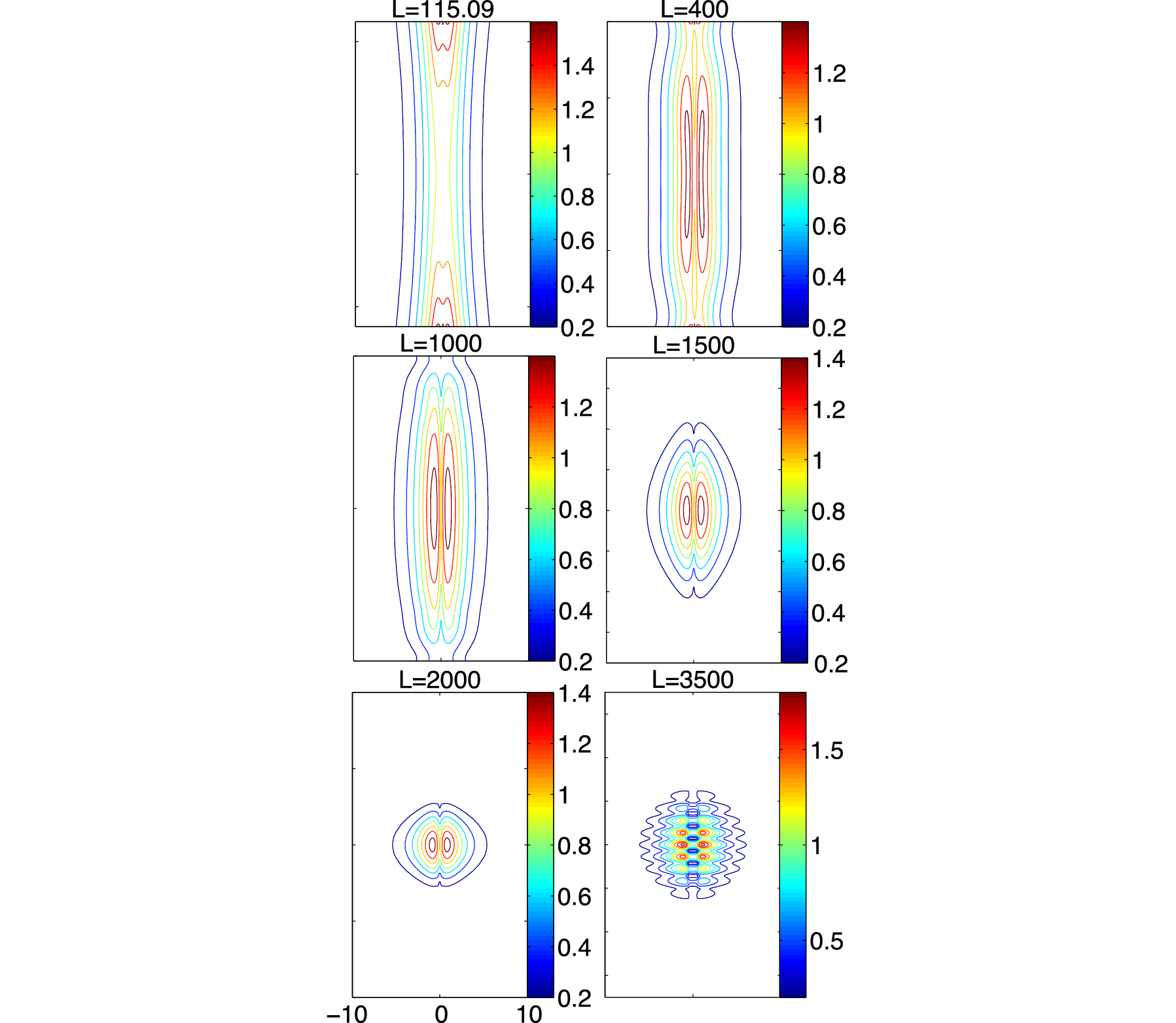}
\par\end{centering}

\caption{(Color online) The eigenfunction$\left|\tilde{\psi}\right|$ for various
$L$. $\eta=10^{-5}$.\label{fig:psi_eigen}}

\end{figure}

Fig. \ref{fig:phi_eigen} and Fig. \ref{fig:psi_eigen} show the eigenfunctions
$\left|\tilde{\phi}\right|$ and $\left|\tilde{\psi}\right|$ for
various $L$, with $\eta=10^{-5}$. The eigenfunction is normalized
such that $\tilde{\psi}(0,0)=1$. The eigenfunction is very smooth
at the critical length $L_{c}=115.09$. As $L$ becomes longer, the
internal layer at $x=0$ become steeper and steeper. Up to $L=1000$,
the eigenfunction still largely covers the whole length along $z$.
At $L=1500$ and $2000$, the eigenfunction becomes localized to the
midplane. This localization is more pronounced in $\tilde{\psi}$.
At $L=3500$, the growthrate is greater than the periodic one at $k_{z}=0$.
The internal layer {}``splits'' into two, and there are wiggles
along the $z$ direction. This is due to the peculiar two-bump structure
in the dispersion relation of the periodic system (Fig. \ref{fig:gamma_vs_kz}).
In the periodic system, $\gamma>\gamma|_{k_{z}=0}$ corresponds to
four nonzero $k_{z}$'s, therefore we expect the line-tied mode to
be dominated by $k_{z}\neq0$ behavior. The four periodic modes have
resonant surfaces at four different locations. The four resonant surfaces
form two pairs, one in $x>0$ and the other in $x<0$. These two pairs
of resonant surfaces roughly correspond to the two parts of the split
internal layer. From the discussion in Ref. \citep{EvstatievDF2006,DelzannoEF2007,DelzannoF2008},
we expect an expansion from four eigenmodes of $\mathcal{L}_{\gamma}$
to be a good approximation when the two resonant surfaces within each
pair become sufficiently close to each other. However, it is found
that for this to occur one would have to go to an enormously large
$L$, and we have trouble achieving full convergence in this regime. 

\begin{figure}
\begin{centering}
\includegraphics[bb=230bp 0bp 640bp 777bp,clip,scale=0.55]{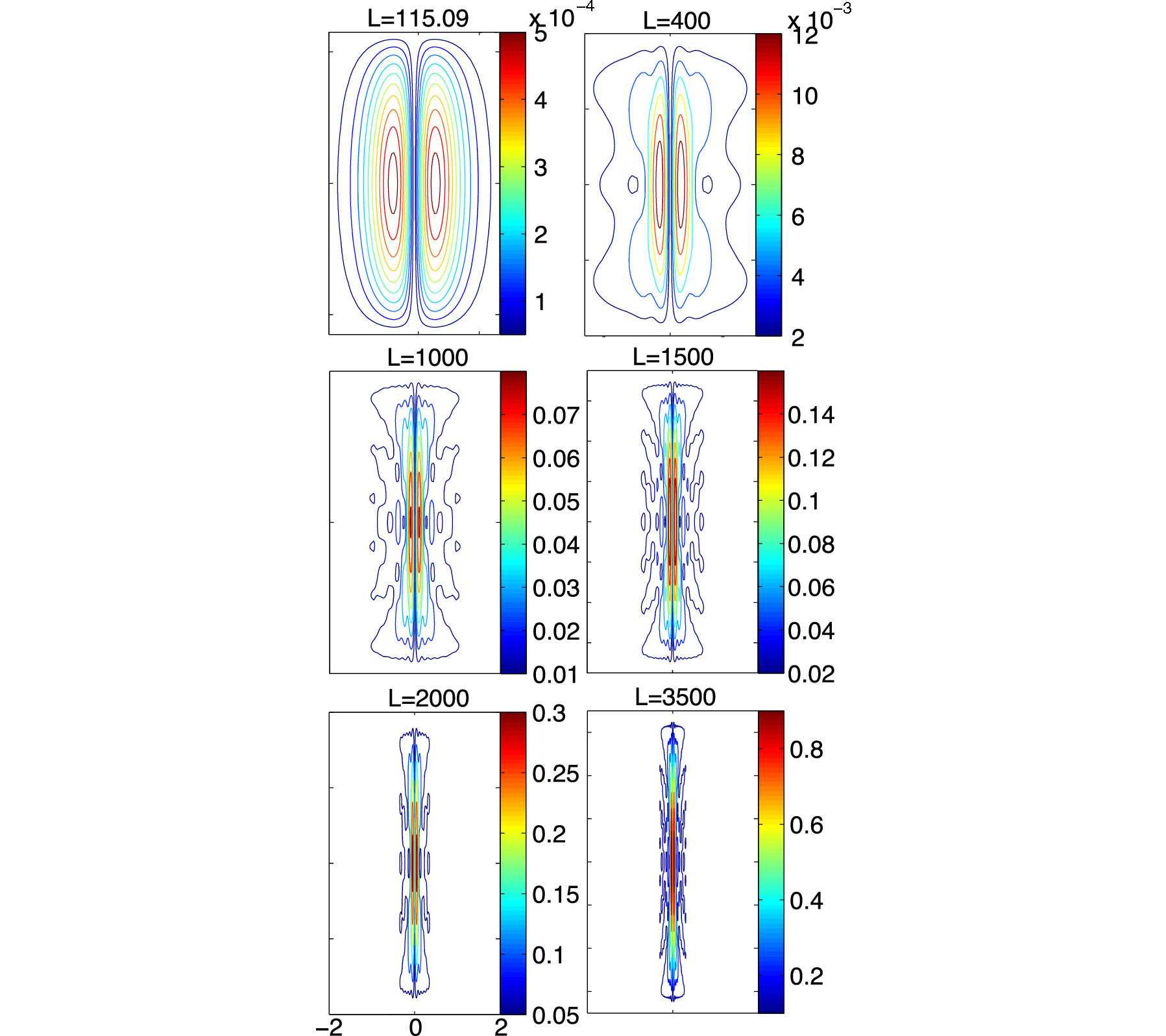}
\par\end{centering}

\caption{(Color online) The eigenfunction$\left|\tilde{\phi}\right|$ for various
$L$ in the force-free approximation. $\eta=10^{-5}$.\label{fig:phi_eigen_ff}}

\end{figure}

\begin{figure}
\begin{centering}
\includegraphics[scale=0.45]{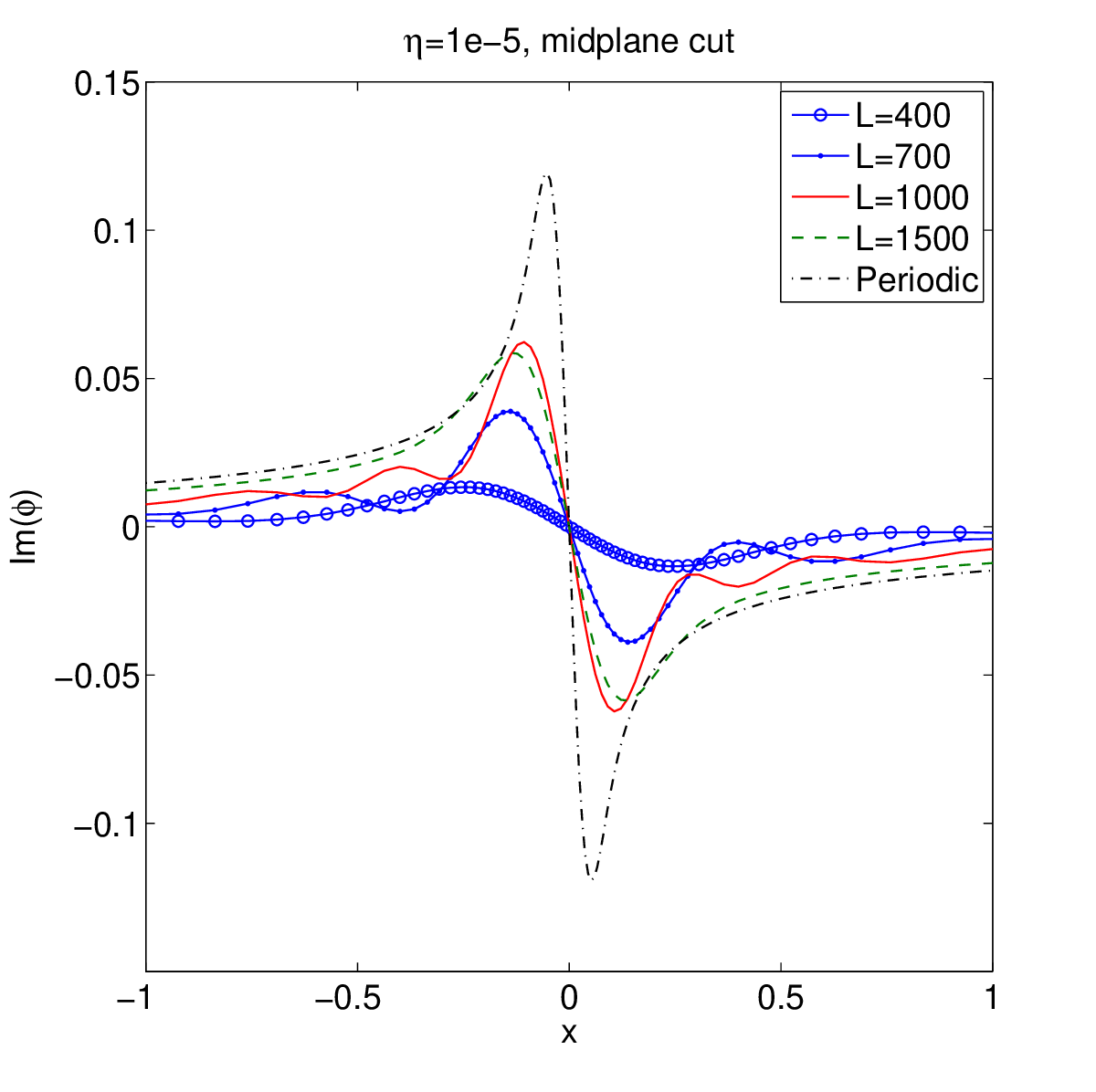}
\par\end{centering}

\caption{(Color online) The midplane cut of $\mbox{Im}(\tilde{\phi})$. Also
shown is the periodic solution with $k_{z}=0$ for reference. \label{fig:phicut}}

\end{figure}

\begin{figure}
\begin{centering}
\includegraphics[scale=0.45]{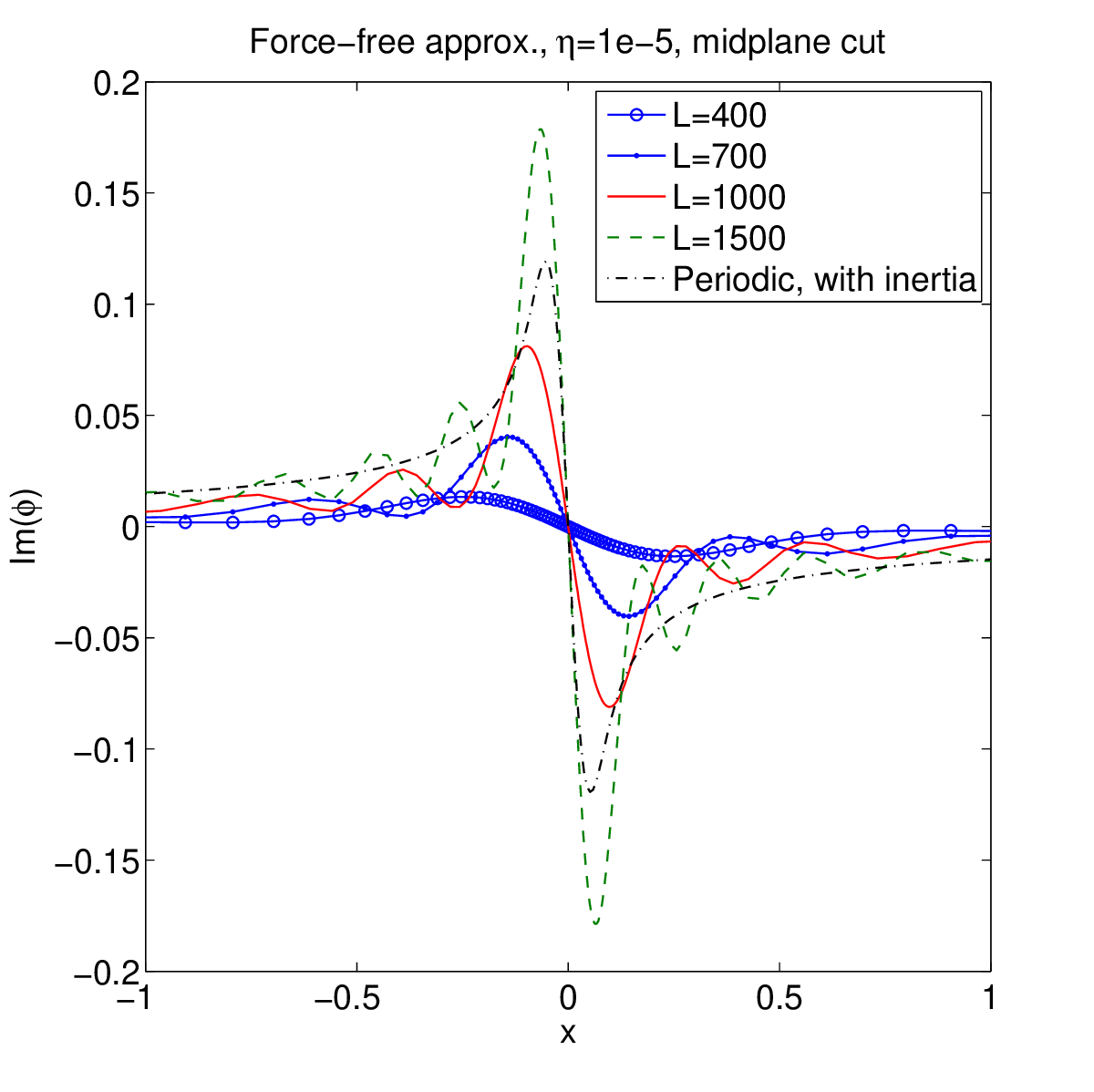}
\par\end{centering}

\caption{(Color online) The midplane cut of $\mbox{Im}(\tilde{\phi})$ for
the force-free approximation. Also shown is the $k_{z}=0$ periodic
solution (with inertia) for reference. \label{fig:phicut_ff}}

\end{figure}

As a comparison, Fig. \ref{fig:phi_eigen_ff} shows the eigenfunction
$\left|\tilde{\phi}\right|$ with the same parameters as Fig. \ref{fig:phi_eigen},
but in the force-free approximation. The force-free approximation
for $L=400$ is almost identical with the real solution. Even at $L=1000$,
the force-free solution is still not too far from the real one. At
$L=1500$ and beyond, the difference between the two is significant.
The same conclusion can be drawn from Fig. \ref{fig:gamma_vs_L},
where the force-free dispersion relation is good for up to $L=1000$,
then it starts to depart from the real one. We also observe that the
internal layer of the force-free eigenfunction keep getting steeper
as $L$ increases. Fig. \ref{fig:phicut} shows the midplane cut of
$\mbox{Im}(\tilde{\phi})$ for various $L$, and Fig. \ref{fig:phicut_ff}
shows its force-free counterpart. Also shown in both figures is the
$k_{z}=0$ periodic eigenmode with the inertia term included, for
reference. Notice that in periodic systems, the inertia term has to
be included to resolve the inner layer,\citep{FurthKR1963,White1986}
while in line-tied configurations the inner layer can be resolved
with the help of the boundary condition (no $\mathbf{k}\cdot\mathbf{B}=0$
singularity) alone. Following the convention of Ref. \citep{DelzannoF2008},
the internal layer width of the force-free approximation is called
the {}``geometric'' width, and that of the periodic eigenmode is
called the {}``tearing'' width. At $L=1500$, the geometric width
is approximately the same as the tearing width, and this is exactly
when the force-free approximation is about to fail. 

Therefore, we come to the conclusion that the transition from the
resistive scaling $\gamma\sim\eta$ to the tearing scaling $\gamma\sim\eta^{3/5}$
occurs when the inertia term is no longer negligible, or equivalently,
when the tearing width is comparable to the geometric width. This
prompts us to do a scaling analysis similar to the one of Ref. \citep{DelzannoF2008}.
Assuming the constant $\psi$ approximation, and letting $\delta$
be the thickness of the internal layer, we have $\mathcal{D}\tilde{\phi}\sim\tilde{\phi}/\delta^{2}$,
$\mathcal{D}\tilde{\psi}\sim\tilde{\psi}\Delta'/\delta$, and $B_{y}\sim B_{y}'\delta$
within the layer, where $\Delta'$ is the jump of $\tilde{\psi}'$
across the layer divided by $\tilde{\psi}$.\citep{FurthKR1963,White1986}
For the periodic case, we have $\gamma\mathcal{D}\tilde{\phi}\sim k_{y}B_{y}\mathcal{D}\tilde{\psi}$
and $\gamma\tilde{\psi}\sim k_{y}B_{y}\tilde{\phi}\sim\eta\mathcal{D}\tilde{\psi}$
within the layer. That is \begin{equation}
\gamma\tilde{\phi}/\delta^{2}\sim k_{y}B_{y}'\Delta'\tilde{\psi},\label{eq:periodic1}\end{equation}
\begin{equation}
\gamma\tilde{\psi}\sim k_{y}B_{y}'\delta\tilde{\phi}\sim\eta\tilde{\psi}\Delta'/\delta.\label{eq:periodic2}\end{equation}
From that, we find the tearing growthrate \begin{equation}
\gamma_{tearing}\sim(k_{y}B_{y}')^{2/5}\Delta'^{4/5}\eta^{3/5},\label{eq:gamma_tearing}\end{equation}
and the tearing width \begin{equation}
\delta_{tearing}\sim\eta^{2/5}\Delta'^{1/5}(k_{y}B_{y}')^{-2/5}.\label{eq:delta_tearing}\end{equation}
For the line-tied case with the force-free approximation, we may assume
$\partial_{z}\sim1/L$ since the eigenfunction spreads over the entire
length (Fig. \ref{fig:phi_eigen_ff}). Within the internal layer,
we have $\partial_{z}\sim k_{y}B_{y}$; i.e., the geometric width\begin{equation}
\delta_{geometric}\sim1/k_{y}B_{y}'L.\label{eq:delta_geo}\end{equation}
From the induction equation, we have \begin{equation}
\gamma_{force-free}\sim\eta\Delta'/\delta\sim\eta k_{y}B_{y}'L\Delta'.\label{eq:gamma_ff}\end{equation}
Also, we may estimate $\tilde{\phi}$ from $\gamma\tilde{\psi}\sim k_{y}B_{y}'\delta\tilde{\phi}$
as \begin{equation}
\tilde{\phi}\sim\eta k_{y}B_{y}'L^{2}\Delta'\tilde{\psi}.\label{eq:phi_est}\end{equation}
The force-free approximation is self-consistent if $\gamma\mathcal{D}\tilde{\phi}\ll\partial_{z}\mathcal{D}\tilde{\psi}$;
i.e., $\gamma\tilde{\phi}/\delta^{2}\ll\tilde{\psi}\Delta'/\delta L$.
Using Eqs. (\ref{eq:delta_geo}), (\ref{eq:gamma_ff}), and (\ref{eq:phi_est}),
the self-consistency condition can be written as \begin{equation}
L\ll L_{t}\sim\eta^{-2/5}\Delta'^{-1/5}(k_{y}B_{y}')^{-3/5},\label{eq:Lt}\end{equation}
where $L_{t}$ is the transition length from resistive scaling to
tearing scaling. The same condition can be obtained if we require
$\delta_{tearing}\ll\delta_{geometric}$.

\begin{figure}
\begin{centering}
\includegraphics[scale=0.45]{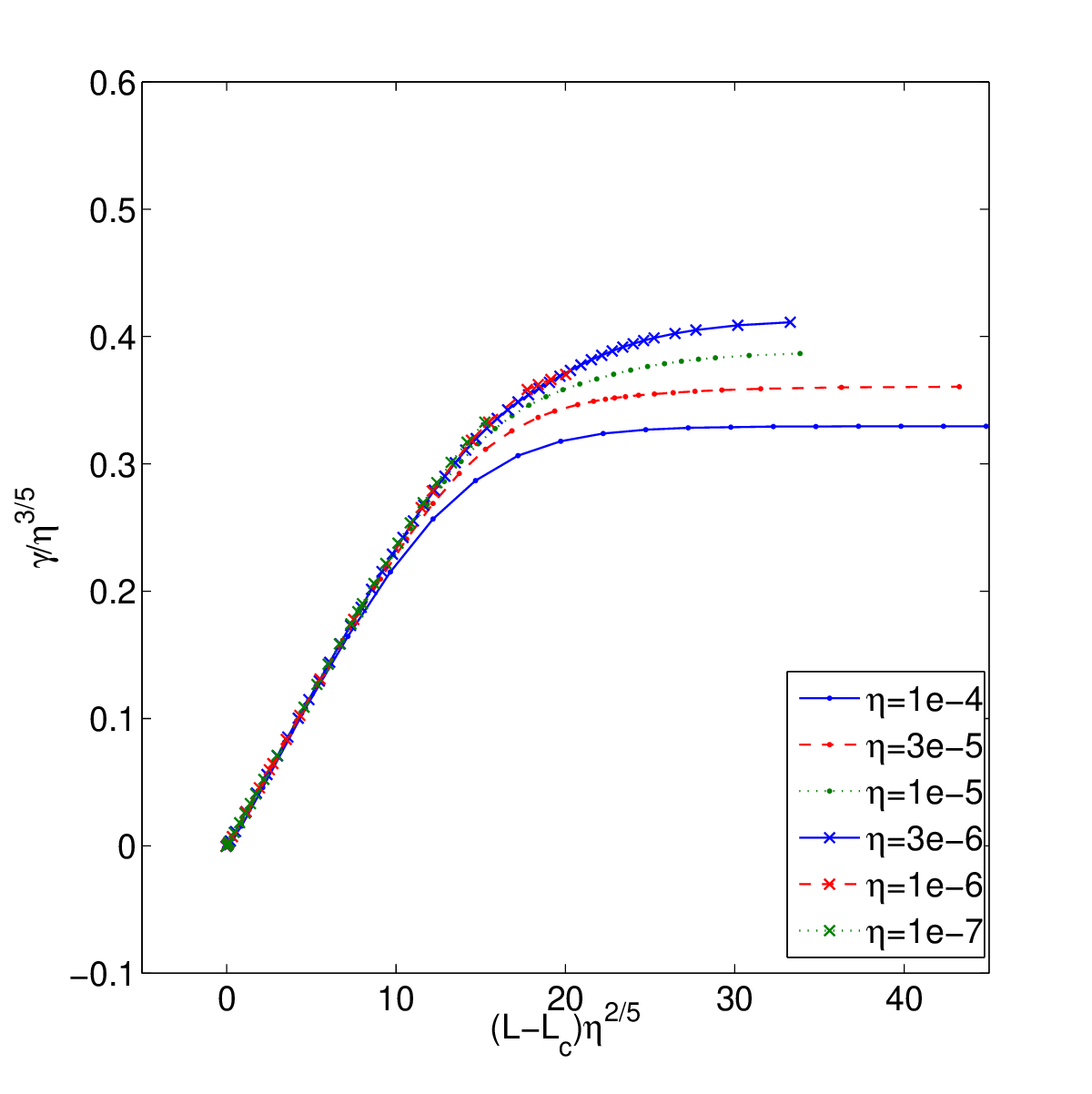}
\par\end{centering}

\caption{(Color online) Rescaled line-tied growthrate as a function of $L$.
\label{fig:Rescaled-line-tied-dispersion}}

\end{figure}

\begin{figure}
\begin{centering}
\includegraphics[scale=0.45]{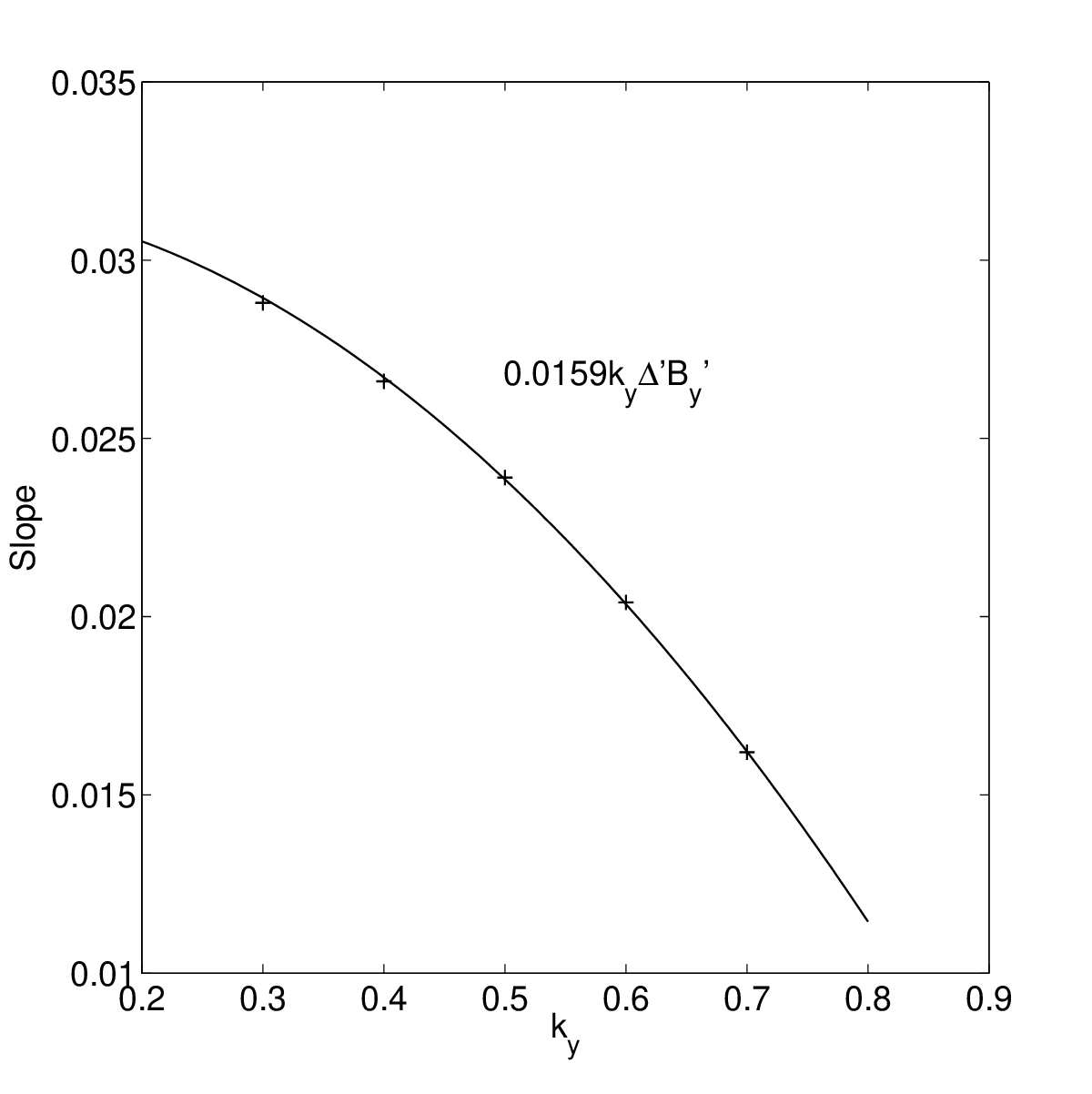}
\par\end{centering}

\caption{Slopes of the force-free $\gamma/\eta$ versus $L$ curve, which is
approximately a straight line, for different $k_{y}$. The curve $0.0159k_{y}\Delta'B_{y}'$
is an excellent fit with the numerical results. \label{fig:slope}}
 
\end{figure}

The scaling analysis predicts that the transition length $L_{t}\sim\eta^{-2/5}$.
Also we know that as $L$ becomes large, $\gamma\sim\eta^{3/5}$.
This motivates us to rescale Fig. \ref{fig:gamma_vs_L} in the following
way. We plot $\gamma/\eta^{3/5}$ versus $(L-L_{c})\eta^{2/5}$. Normalizing
$\gamma$ by $\eta^{3/5}$ brings the asymptotic regime to the same
level, while multiplying $(L-L_{c})$ by $\eta^{2/5}$ brings the
transition length to the same place. The result is shown in Fig. \ref{fig:Rescaled-line-tied-dispersion}.
Clearly those curves for different $\eta$ become close to each other,
especially those with smaller $\eta$ are almost identical. Therefore
the numerical results agree with the $L_{t}\sim\eta^{-2/5}$ prediction.
The analysis also predicts $\gamma_{force-free}\sim\eta k_{y}B_{y}'L\Delta'$.
We can see from Fig \ref{fig:gamma_vs_L} that the force-free dispersion
relation is very close to a straight line. That suggests that we may
associate the slope with $\gamma_{force-free}/\eta L\sim k_{y}B_{y}'\Delta'.$
For the Harris sheet profile, $\Delta'=2(1/k_{y}-k_{y})$. To test
the conjecture we have to find the slope for different $k_{y}$. The
slope and the critical length for different $k_{y}$ are summarized
in Table \ref{tab:Dependence-on-k}. We found that $L_{c}$ only weakly
depends on $k_{y}$. Fitting the slope with $\mbox{const}\times k_{y}\Delta'B_{y}'$
gives $\mbox{slope}\simeq0.0159k_{y}\Delta'B_{y}'.$ This turns out
to be an excellent fit, as shown in Fig. \ref{fig:slope}.

\begin{table}[t]
\begin{centering}
\begin{longtable}{cccc}
\hline
\hline 
$k_{y}$ & $L_{c}$ & Slope & $k_{y}\Delta'B_{y}'$\tabularnewline
\hline 
0.3 & 112.85 & 0.0288 & 0.182\tabularnewline
\endhead
0.7 & 141.09 & 0.0162 & 0.102\tabularnewline
\hline
\hline
\endfoot
0.4 & 111.50 & 0.0266 & 0.168\tabularnewline
0.5 & 115.09 & 0.0239 & 0.150\tabularnewline
0.6 & 123.86 & 0.0204 & 0.128\tabularnewline
\end{longtable}
\par\end{centering}

\caption{ The critical length $L_{c}$ and the slope of the force-free dispersion
relation $\gamma/\eta$ versus $L$ for different $k_{y}$. Also shown
is the corresponding $k_{y}\Delta'B_{y}'$. The scaling analysis suggests
that slope $\sim k_{y}\Delta'B_{y}'$. \label{tab:Dependence-on-k}}

\end{table}

\section{Summary and Discussion\label{sec:Summary-and-Discussions}}

In this work, we study the effects of line-tying on the resistive
tearing instability in great detail. We found that line-tying has
a stabilizing effect, and instability occurs only when the system
length is greater than a critical length $L_{c}$, which is independent
of the resistivity $\eta$. When the system length is not too long,
the plasma inertia is negligible. In other words, the plasma is approximately
force-free throughout the whole domain. In this regime, the growthrate
$\gamma$ is proportional to $\eta$ and the internal layer of the
unstable mode is much wider compared to the periodic mode. The tearing
mode scaling $\gamma\sim\eta^{3/5}$ is recovered when the system
becomes sufficiently long. The transition from $\gamma\sim\eta$ to
$\gamma\sim\eta^{3/5}$ occurs when the plasma inertia is no longer
negligible within the resistive layer; this is also when the {}``geometric''
width becomes comparable to the {}``tearing'' width. The transition
length $L_{t}$ scales like $\eta^{-2/5}$. 

Physically it is not hard to understand why the growthrate in the
force-free approximation is proportional to $\eta$. In the original
system, there are two time scales, the Alfv\'enic time scale and
the resistive time scale; and the tearing mode grows at a hybrid time
scale of the two. If the inertia is neglected completely, the resistive
time scale is the only one left; therefore the mode can only grow
on the resistive time scale. However, neglecting the inertia in the
entire domain is possible only in a line-tied system, as the line-tied
boundary conditions serve as a means of resolving the singularity
at the rational surface. In a periodic system, the inertia term always
comes into play in the vicinity of the rational surface.

The periodic growthrate and the growthrate from the force-free approximation
may be regarded as two asymptotes of the line-tied growthrate. The
force-free growthrate $\gamma_{force-free}$ is approximately linear
with respect to $L$ and the slope is found to be related to $\Delta'$.
To a very good approximation $\gamma_{force-free}\simeq0.0159k_{y}\Delta'B_{y}'\eta(L-L_{c})$,
where $B_{y}'$ is evaluated at $x=0$. Unfortunately, we do not have
a good way to estimate the critical length $L_{c}$. 

To a large extent, our findings are in agreement with the earlier
study by Delzanno and Finn in cylindrical geometry.\citep{DelzannoEF2007}
One major discrepancy is that we found $L_{t}\sim\eta^{-2/5}$ while
they reported $L_{t}\sim\eta^{-1/4.2}\sim\eta^{-0.24}$ (They actually
consider the crossover resistivity $\eta_{cross}$ for a given $L$,
and find $\eta_{cross}\sim L^{-4.2}$; but this is just a matter of
which variable is held fixed). A possible explanation for the discrepancy
is that they use a fairly large viscosity, which remains unchanged
when other parameters are varied. We have not studied the effect of
viscosity in great detail. Some preliminary results indicate that
viscosity does have a significant effect, and $L_{t}\sim\eta^{-2/5}$
scaling is no longer valid when a large viscosity is added. 

We conclude with a brief discussion of our findings in the context
of solar corona. The solar corona is very highly conducting, and the
aspect ratios of observed coronal loops are $10$ -- $100$, far smaller
than necessary for growthrates in the tearing regime. Therefore it
is likely the resistive instability will operate in the $\gamma\sim\eta$
regime. An immediate question is that, since the resistive time scale
is so long in the solar corona, is the resistive instability relevant
at all? We do not have a clear answer as yet. \textcolor{black}{It
is quite clear that our model is highly idealized and has some obvious
shortcomings. An immedaite concern is that footpoint drivers are not
included in our analysis; therefore the equilibrium profile will decay
on a resistive time scale $\tau_{\eta}$. Our analysis implicitly
assumes either that $\gamma\gg1/\tau_{\eta}$, such that the resistive
decay may be neglected, or that the equilibrium is maintained by some
external current drive, as in laboratory devices. In the solar corona
the electric current is driven by footpoint motions, and the assumption
is clearly inadequate when $L_{c}<L\ll L_{t}$, as the growthrate
$\gamma$ is either comparable to or even smaller than $1/\tau_{\eta}$.
A remedy to this problem would be adding a footpoint driver to the
system to maintain the equilibrium. This will add an overall background
flow to the equilibrium and potentially have some effect on the stability,
especially when the system is close to marginal stability. Detailed
analysis is left to a future study. Furthermore, the rigid wall boundary
condition, which we take for simplicity, may be too restrictive as
argued by some workers. This is especially true when considering modes
with $\gamma\sim\eta$, as the photosphere may be able to respond
during the slow time scale of growth. A proper treatment requires
some detailed knowledge of the photosphere interior. \citep{Hassam1990,Hood1986,VanDerLindenHG1994,Zweibel1985}
In light of these shortcomings, our findings certainly need furthur
scrutiny when applied to the solar corona.}

As a final remark, the nonlinear evolution of the mode has to be studied.
Indeed, even in the classical tearing mode theory without line-tying,
the unstable mode evolves into the nonlinear regime very quickly.\citep{Rutherford1973}
Our preliminary study indicates that thin current filaments can be
created within the original current layer as the mode evolves nonlinearly.
Therefore the role resistive instabilities play in current sheet thinning
cannot be overlooked. Another interesting point is that, since the
stability criterion does not depend on the resistivity but the growthrate
does, the solar corona may appear to be quiescent but already be in
an unstable configuration. The slow growthrate of the resistive instability
allows the corona to be driven well beyond the critical point, therefore
more free energy could be stored. Yet the resistive instability, however
slow it may be, also provides a means for the corona to tap into the
free energy in an otherwise ideally stable configuration. The multiple
roles the resistive instability may play here leave much to ponder
about. We hope to address some of these issues in the future. 

\begin{acknowledgments}
It is a pleasure to acknowledge beneficial discussions with Dr. G.
L. Delzanno, Dr. E. G. Evstatiev, and Dr. J. M. Finn about their numerical
methods and the physics of line-tying. Yi-Min Huang would like to
thank Daniel Lecoanet and Nick Murphy for inspirations. Yi-Min Huang
also thanks Prof. A. Bhattacharjee for the support to present this
work in the annual meeting of American Physical Society, Division
of Plasma Physics. The authors gratefully acknowledge the insightful
comments from the referee. This research is supported by the National
Science Foundation, Grant No. PHY-0215581 (PFC: Center for Magnetic
Self-Organization in Laboratory and Astrophysical Plasmas).
\end{acknowledgments}
\bibliographystyle{apsrev}

\newpage{}

\end{document}